\documentclass[12pt]{article}
\usepackage{amssymb}
\usepackage{graphicx}
\usepackage{amsmath}
\newtheorem{theorem}{Theorem}

\newtheorem{corollary}[theorem]{Corollary}

\newtheorem{lemma}[theorem]{Lemma}

\newtheorem{proposition}[theorem]{Proposition}
\newtheorem{remark}[theorem]{Remark}

\newenvironment{proof}[1][Proof]{\textbf{#1.} }{\ \rule{0.5em}{0.5em}}

\usepackage{setspace}
\title{Two Families of Quantum Codes Derived from Cyclic Codes } 
\author{K. Guenda}
\date{}
 
\begin{document}
\maketitle 
\begin{center}Faculty of Mathematics, University of Sciences and
Technology of Algiers \\
B. P 32 El Alia 16111 Bab Ezzouar, Algiers, Algeria\\guendakenza@hotmail.com \end{center}

\begin{abstract}
We characterize the  affine-invariant maximal extended cyclic
codes. Then by the $CSS$ construction,
we derive from these codes a family of pure quantum
codes. Also for $\textrm{ord}_nq$ even, a new family of degenerate quantum
stabilizer $[[n,1, \geq \sqrt{n}]]_q$ codes
  is derived from the classical duadic codes. This answer an open problem asked by Aly et al.
\end{abstract}
\bigskip

\section{Introduction}
In this paper we consider maximal-affine invariant codes and duadic
codes. The first are extended cyclic codes and the second are
cyclic codes. We characterize the form of the maximal-affine invariant
codes and the cases when they
 contain their dual codes. Then by using the $CSS$ construction we derive
 a new family of quantum codes. For $q$ a prime number, the obtained 
 quantum codes are pure.
In the second part of this paper, the interesting relation between the duals of the pairwise of
even-like and odd-like duadic codes allow us to give a $[[n,1,
    \geq d]]_q$ family of quantum degenerate codes. The degeneracy of
these codes is an interesting propriety as it was shown in \cite{Aly}.
 The construction given here is more general than in \cite{Aly}, since
 it does not require that $\textrm{ord}_nq$ is odd.
 Note that the quantum codes are used to protect quantum information
 over noisy quantum channels.

\section{The Maximal Affine-Invariant Codes}

Let $q=p^r$ be a power of a prime number $p$, $K=\mathbb{F}_{p^r}$ and $G=\mathbb{F}_{p^m}$ be respectively the
 finite field  with $p^r$ and $p^m$ elements and $m=rm'$. A primitive
 cyclic code
 $C$ of length $n=p^m-1$ is an ideal of the algebra
$R=\mathbb{F}_q[x]/(x^n-1)$   
 generated by a polynomial $g(x)$ which divides $x^n-1$, and is uniquely determined by its
 defining set 
 $T=\{0 \leq i \leq 1 | \ g({\beta}^i)=0\},$ where $\beta$ is an
 $n^{th}$ primitive root of the unity in $G$. The set 
  $T$ is then an union of cyclotomic classes $Cl(j)=\{jq^{l}\mod n\,|\,l\in
 \mathbb{Z}\}$. If $T$ contains one
 cyclotomic class, $C$ is called maximal
 cyclic code.
We associate with the code $C$ the code $$\Hat{C}=\{ (x_0, \ldots,x_n) \,|\,  (x_1,
 \ldots,x_n) \in C \text{ with } x_0=-\sum_{i=1}^nx_i  \},$$
called the extended code of $C$. The defining set $T$ of $C$ is also
 called the defining set of $\Hat{C}$. 
Since the code $\Hat{C}$ is of length $p^m$, the field $G$ is viewed as
 the support field and the coordinate position of the codewords are
 labelled by the elements of $G$, hence can be considered as a subspace
 of the group algebra $A=K[\{G,+\}]$, which is the group algebra of the
 additive group of $G$ over $K$, 
$$A=\{\sum \limits_{g \in G}x_gX^g, x_g \in K\}.$$The operations in
 $A$ are given by 
$$a\sum \limits_{g \in G}x_gX^g+b\sum \limits_{g \in G}y_gX^g=\sum
 \limits_{g \in G}(ax_g+by_g)X^g,\, a, b \in K,$$
and
$$\sum \limits_{g \in G}x_gX^g\times \sum \limits_{g \in G}y_gX^g=\sum
\limits_{g \in G}(\sum \limits_{h+k=g }x_hy_k)X^g.$$The zero and unity
of $A$ are $\sum \limits_{g \in G}0X^g \text{ and } X^0.$

The permutation group $Per(C)$ of a linear code code of length $n$ is
  the subgroup of $Sym(n)$ which leaves the code $C$ invariant. An
  extended cyclic code $\Hat{ C }$ of length $p^m$ is said to be affine-invariant provided
  its permutation group contains
$AGL(1,p^m)
=\{ \sigma \,|\, f_{\sigma}(X) =aX+b, a \in G^*, b \in G \}$. 

A partial order on $S=[0,p^m -1[$ is defined by:\\
If $s= \sum \limits_{i=0}^{m-1}s_ip^i,$
$t= \sum \limits_{i=0}^{m-1}t_ip^i$ are the $p$-adic expansions of $s$
and $t$, then $$s \prec t \iff s_i \leq t_i, \forall i =0,
\ldots, m-1, s_i \in [0,p-1].$$
The following well known result $[3,6]$ characterizes the affine-invariant
  codes by their defining sets.
\begin{theorem}
\label{kasami}
An extended cyclic code $\Hat{ C }$ with defining set $T$ is 
 affine-invariant if and only if $T$ satisfies
$$ \forall t \in T , \forall s \in S, s \prec t \Rightarrow s \in T.$$
\end{theorem}
As a consequence we get the following result.
\begin{proposition}
\label{main}
A maximal extended cyclic code of $A$ is affine-invariant if and only if $$T=Cl(p^j).$$ 
\end{proposition}
\begin{proof}
Let $T$ be the defining set of a maximal extended cyclic code, it is clear that
if $T =Cl(p^j)$ with $0\leq j \leq m-1$, then for $s\in S$ and $t\in T$ we can have $s \prec t$
only for $s=t$. Therefore by Theorem \ref{kasami} the set $T$ is the defining
set of an affine-invariant code.
 
Let $T=Cl(t)$ be a cyclotomic class modulo $p^m-1$, where $t$ is the
smallest element for the order $ (\leq)$. For $t \in ] p^j, p^{j+1}[$, 
the $p$-adic expansion of $t$ is $$t = \sum \limits_{i=0}^{j-1} s_i
    p^i + \alpha p^j,$$
with $0 \leq s_i <p$ and $1 \leq \alpha $. Since $1\leq \alpha $, then
$p^j \prec t$, and by Theorem \ref{kasami} the extended code associated to $T$ cannot be affine-invariant.
\end{proof}
\begin{remark}
\label{number}
Using the fact that  $\textrm{ ord }_{p^m-1} p^r=\frac{m}{r}$, we can
obviously deduce that the cardinality of $Cl(p^j)$ is
$\frac{m}{r}$. Hence for fixed parameters, the number of distinct maximal affine-invariant
codes  is $r$.
\end{remark}

\section{Quantum Affine-Invariant Codes}
From the classical affine-invariant codes of the previous section we
can directly obtain a family of quantum codes by using the called $CSS$
given by the following Lemma.
\begin{lemma}
\label{day}
Let $C_1 = [n, k_1 , d_1 ]$ and $C_2 = [n, k_2 , d_2 ]$
be linear codes over $\mathbb{F}_q$ with $C_2^{\bot} \subset C_1$. Furthermore, let
                                
$$d_Q = \min \{ wt (v): v \in (C_1 \setminus C_2^{\bot} ) \cup (C_2 \setminus C_1^{\bot}
  )\} \geq \min (d_1,d_2).$$ Then there exists
a $[[n, k_1 + k_2-n , d_Q]]_q$ quantum code.
\end{lemma}
\begin{proof}
See for instance \cite{Grassl}.
\end{proof}\\

By Lemma \ref{day}, if a linear $[n,k,d]$ code $C$ contains its dual
$C^{\bot}$, then there exists a quantum code with parameters $$[[n,2k-n,
   d_Q \geq d]]_q,$$that is pure up to $d$. Now we prove the following Lemma which
tell us when a maximal affine-invariant code can contains its dual.
\begin{lemma}
\label{awal}
Let $\Hat{C}$ be an extended maximal affine-invariant code
$[p^m,p^m-1-\frac{m}{r},d]$, then if $p >3$ or
$m>2$ or
$r\neq 1$ we have $\Hat{C}^{\bot} \subset \Hat{C}$. 
\end{lemma}  
\begin{proof}
Let $T^{\bot}$ be the defining set of $\Hat{C}^{\bot}$, we have the
  relation  $T^{\bot}=\{s \in Z_n \,|\, n-s \notin T \}$. To have
  $\Hat{C}^{\bot} \subset \Hat{C}$ it suffices to have
$T \subset T^{\bot}$, which is then equivalent to the following
  statement 
\begin{equation}
\label{djoh}
s \in T \Rightarrow p^m-1-s \notin T.
\end{equation}
Let $T=Cl(p^j)$, for simplicity we consider $s=p^j$; the general  case
is therefore obviously deduced. Seeking a contradiction we assume that $p^m-1-p^j \in
T$, then it exists $0 \leq \alpha \leq \frac{m}{r}-1$ such that
$p^m-1-p^j=p^{\alpha r+j} \mod (p^m-1)$, this is equivalent to have $ p^j(1+p^{\alpha
  r})=k(p^m-1)$ for some $k \geq 1$. Hence $1+p^{\alpha r}= k'(p^m-1)$
for some $k' \geq 1$, this implies that
\begin{equation}
\label{djas}
1+p^{\alpha r} \geq p^m-1.
\end{equation}
Since $\alpha \leq \frac{m}{r}-1$, then (\ref{djas}) gives 
 $2 \geq p^{m-r}(p^r-1)$, this is possible only for ($r=m=1, p \leq 3$)
 or ($r=1, m=2, p=2$). Therefore under the hypothesis $p >3$ or
$m>2$ or
$r\neq 1$ we have $\Hat{C}^{\bot} \subset \Hat{C}$.
\end{proof}

From Lemma \ref{awal} and the $CSS$ construction we get the
following result.
\begin{theorem}
\label{pat}
Let $q=p^r$ with $p$ a prime number, $m$ a positive integer and
$n=p^m-1$. If $p >3$ or
$m>2$ or
$r\neq 1$, then there exists a quantum code with parameters
$$[[p^m,p^m-2-2\frac{m}{r},  \frac{m}{r}+2 \geq d_Q \geq d_A]]_q,$$
where $d_A$ is the minimum
distance of an extended maximal affine-invariant code.
\end{theorem}
\begin{proof}
Proposition \ref{main} and Lemma \ref{awal} imply the existence of affine-
invariant code $\Hat{C}$ which contains its dual with
parameters
$[p^m,p^m-1-\frac{m}{r},d_A]$.
 Therefore by Lemma \ref{day}, we have the
existence of a $[[p^m,p^m-2-2\frac{m}{r},d_Q\geq d_A]]_q$ quantum
code. The quantum Singleton bound gives : $d_Q \leq 2+\frac{m}{r}$.
\end{proof}
\begin{corollary}
Assuming  that $r=1$ and ($p > 3$ or $m
>2$), there exists a pure quantum code with parameters $$[[p^m,p^m-2-2m,  m+2 \geq d_Q= d_A]]_p,$$
where $d_A$ is the minimum distance of an extended maximal affine-invariant code.
\end{corollary}
\begin{proof}
Assume $r=1$ and ($p > 3$ or $m
>2$), then Theorem \ref{pat} gives the existence of a quantum code with
parameters $[[p^m,p^m-2-2m,  m+2 \geq d_Q \geq d_A]]_p$. 
Form Remark \ref{number} there is a unique maximal affine invariant
code $\Hat{C}$, which is the extended $BCH$ code with designed distance 2.
Then from the Weil bound given in \cite{wolfmann}
the minimum distance of $\Hat{C}^{\bot}$ is $d = p^m -p^{m-1}$ or $d =
p^m -p^{m-1}+1$ . Hence from Lemma \ref{day}
the quantum code is pure;
   $d_Q=d_A$.
\end{proof}
\section{Duadic Codes}
Throughout this part we denote by $n$ an odd integer and by $q$ a prime
power. The notation $q \equiv \square \mod n$ express the fact that $q$ is a
quadratic residue modulo $n$. We write $p^{\alpha} \mid \mid m$ if
and only if the integer $m$ is divisible by $p^{\alpha}$ but not by
$p^{\alpha +1}$. For an integer $a$ such that $(a,n)=1$, $\mu _a:i \longmapsto ia \mod n$ denote
a permutation on $Z_n =\{0,\ldots, n-1\}$.

\subsection{Classical Duadic Codes}
Let $q$ be a prime power and $n$ an odd integer
such that $(n,q) =1$. 
Let $S_1$ and $S_2$ be unions of cyclotomic cosets modulo $n$ such that
  $S_1 \cap S_2 = \emptyset$,
 $S_1 \cup S_2 = Z_n \setminus \{0\}$ and
 $aS_i  \mod n = S_{(i+1) \mod 2}$. Then the triple $\mu _a, S_1, S_2$
is called splitting modulo $n$. The odd-like duadic codes $D_1$ and
  $D_2$ are the cyclic codes over $\mathbb{F}_q$ with defining
  set respectively
  $S_1$ and $S_2$. The even-like duadic codes $C_1$ and $C_2$ are the
  cyclic codes over $\mathbb{F}_q$ with defining
  set respectively
$\{0\}\cup S_1$ and $\{0\} \cup S_2$. 
\begin{theorem}
Duadic codes of length $n$ over $\mathbb{F}_q$ exist if and only if $$q \equiv
\square \mod n.$$
\end{theorem} 
\begin{proof}
This is well-known, see for example \cite{smid}.
\end{proof}
\begin{lemma}
\label{square}(square root bound)\\
Let $D_1$ and $D_2$ be a pair of odd-like duadic codes of length $n$
over $\mathbb{F}_q$. Then they have the same minimum distance say $d$,
and verifies
\begin{enumerate}
\item $d^2 \geq n$ 
\item $d^2 -d +1 \geq n$ if the splitting is given by $\mu _{-1}$
\end{enumerate}
\end{lemma}  
\begin{proof}See [8, Theorem 7].
\end{proof}
\subsection{Quantum Duadic Codes}

If we are over $\mathbb{F}_{q^2}$, we can consider the Hermitian duality. Furthermore we
have the existence of duadic codes over $\mathbb{F}_{q^2}$ for all
$n$, when $(n,q^2)=1$, since $q^2 \equiv \square \mod n$.

The following Lemma gives the same construction as Lemma \ref{day} in the
Hermitian case.
\begin{lemma}
\label{soubh}
If there exists a classical linear $[n,k,d]_{q^2}$ code $C$, such that
  $ C \subset C^{\bot h} $. Then there exists an $[[n, n-2k, \geq d]]_q$
  quantum code that is pure to $d$.
\end{lemma}
\begin{proof} See for instance [4, Corollary 2].
\end{proof}
\begin{lemma}
\label{fadila}
Let $C_i$ and $D_i$ respectively be the even-like and odd-like codes
over $\mathbb{F}_{q^2}$, where $i \in \{0,1\}$. Then $C_i^{\bot h} = D_i$
 if and only if there is a $q^2$-splitting given by $\mu_{-q}$, that is,
 $-qS_i \equiv S_{(i+1 \mod 2)} \mod n.$
\end{lemma} 
\begin{proof} See [9, Theorem 4.4] 
\end{proof}  
   
We prove the following Lemma which gives some properties of
$\textbf{ord}_nq$ which are useful for our construction.
\begin{lemma}
\label{najiya}
Let $q$ be a prime power and $n$ an integer, then :
\begin{enumerate}
\item If $\textbf{ord}_nq$ is odd, then $\textbf{ord}_nq^2=\textbf{ord}_nq$. 
\item If $\textbf{ord}_nq$ is even, then $\textbf{ord}_nq^2=
  \frac{\textbf{ord}_nq}{2}$
\end{enumerate}
\end{lemma}
\begin{proof}
 \begin{enumerate}
\item Let $r=\textbf{ord}_nq$ and $r'=\textbf{ord}_nq^2$, 
then $ r | 2 r'$, hence $ r | r'$ since $r$ is odd. On the other hand, we
have $q^{2r} \equiv 1 \mod n$, therefore $r' | r$, finally
$r=r'$.

\item Let $r=\textbf{ord}_nq$ and $r'=\textbf{ord}_nq^2$, we
  have $q^{2r'} \equiv 1 \mod n$ which implies that $r | 2 r'$.
Since $r$ is even, then $q^{2\frac{r}{2}} \equiv 1 \mod n $ therefore
  $ r' | \frac{r}{2}$, hence $r'=\frac{r}{2}$.
\end{enumerate}
\end{proof}

\begin{proposition}
\label{houda}
Let $n= \prod p_i^{m_i}$ be the prime factorization of the odd integer
$n$, where each $m_i > 0$, such that every $ p_i
\equiv -1 \mod 4$, then $\textbf{ord}_nq^2$ is odd. 
\end{proposition}
\begin{proof}
We have $\textbf{ord}_nq^2 =lcm(\textbf{ord}_{p_i^{m_i}}q^2)$. By
\cite{leveque} we have that
$\textbf{ord}_{p_i^{m_i}}q^2=p_i^{m_i-z_i}t_i$, where
$t_i=\textbf{ord}_{p_i}q^2$ and $z_i$ is such that
$p_i^{z_i}
\mid \mid q^{t_i}-1$. Hence  to prove the proposition it suffices to prove that
$\textbf{ord}_{p_i}q^2$ is odd if $ p_i \equiv -1 \mod 4$.
We have to consider two cases:
\begin{enumerate}
\item $q \equiv \square \mod p_i$, then $\textbf{ord}_{p_i}q$  divides
  $\frac{p_i-1}{2}$, this implies that $\textbf{ord}_{p_i}q$ is odd
  otherwise $ p_i
\equiv 1 \mod 4$. Then from Lemma \ref{najiya} $\textbf{ord}_{p_i}q^2$
  is odd.  
\item $q$ is not a square modulo $p_i$, then
  $\textbf{ord}_{p_i}q=p_i-1$ is even and then from Lemma \ref{najiya}
  $\textbf{ord}_{p_i}q^2=\frac{p_i-1}{2}$. The last quantity must be odd
  otherwise we get $ p_i \equiv 1 \mod 4$. 
\end{enumerate}
\end{proof}
\begin{lemma}
\label{ummy} 
Let $n= \prod p_i^{m_i}$ be an odd integer such that
$\textbf{ord}_{n}q^2$ is odd. Then $\mu_{-q}$ gives a splitting of $n$
over $\mathbb{F}_{q^2}$. Furthermore $\mu_{-1}$ and $\mu_{-q}$ give the
same splitting.
\end{lemma}
\begin{proof}
Let $\{S_1,S_2,a\}$ be a splitting over
$\mathbb{F}_{q^2}$. We have $q^{\textbf{ord}_{n}q^2} S_i =S_i \mod
n$. If $\textbf{ord}_{n}q^2 =2k+1$, then $q^{2k+1}S_i=q^{2k}qS_i=S_i$,
but since we are over $\mathbb{F}_{q^2}$ we have $q^{2k}S_i=S_i$,
hence $qS_i=S_i$. That means that $\mu_{q}$ fixes each $S_i$ if the
multiplicative order of $q^2$ modulo $n$ is odd.
From [9, Lemma 5], there exists a $q^2$-splitting of $n$ given by
$\mu _{-1}$ if and only if and only if $\textbf{ord}_{n}q^2$ is odd. Hence
$-S_i \equiv S_{(i+1 \mod 2)} \mod n$; since  $\mu_{q}$ fixes $S_i$
hence $\mu_{-q}$ gives a
$q^2$-splitting of $n$. Conversely if $\mu _{-q}$ gives a
$q^2$-splitting of of $n$, then $-q S_i \equiv S_{(i+1)\mod 2} \mod
n$. Since $\mu _q$ fixes $S_i$ we have $-qS_i \equiv -S_i \equiv
S_{(i+1 \mod 2)} \mod n $; hence $\mu _{-q}$ gives the same splitting
as $\mu _{-1}$.
\end{proof}

Now by using the results of Lemma \ref{ummy}, we construct a quantum
family codes. This construction is more general than the ones done in
[1, Theorem 10]. 
\begin{theorem}\label{Aya}
Let $n$ be an odd integer such that $\textbf{ord}_{n}q^2$ is odd. Then
there exists an $[[n,1,d]]_q$ quantum code with $d^2 -d+1 \geq n$.
\end{theorem}
\begin{proof}
From Lemma  \ref{ummy}  there exist duadic codes $C_i \subset D_i$
with splitting given by $\mu_{q}$ and $\mu_{-1}$. This means that the
$C_i \subset C_i^{\bot h}= D_i$, by Lemma \ref{fadila}. Therefore by
Lemma \ref{soubh} there exists an
$[[n,n-(n-1),d]]_q$ quantum code with $d=wt(D_i \setminus C_i)$. Since
$\mu _{-1}$ gives a splitting, from Lemma \ref{square} we have
$d^2-d+1 \geq n$. 
\end{proof}

Now by using Theorem \ref{Aya} we construct a family of degenerate
codes.
\begin{theorem}\label{naima}
Let $n= \prod p_i^{m_i}$ be an odd integer such that every $ p_i
\equiv -1 \mod 4$. Let
$t_i=\textbf{ord}_{p_i}q^2$ and $p_i^{z_i} \mid \mid q^{2t_i} -1$. Then for
$m_i>2z_i,$ there exists degenerate $[[n,1,d]]_q$ quantum codes pure
to $d' \leq \min \{p_i ^{z_i}\} <d$ with $d^2-d +1 \geq n$. 
\end{theorem}
 \begin{proof}
From [1, Lemma 5] there exists an even-like duadic codes with
parameters $[n,(n-1)/2,d']_{q^2}$ and $d' \leq \min\{p_i^{z_i}\}$. Then
by [11, Theorem 8] the splitting for these codes is given by $\mu _{-1}$. By Proposition \ref{houda} $\textbf{ord}_{n}q^2$ is odd
and then by Lemma \ref{ummy} $\mu _{-q}$ also gives a splitting for
these codes. Hence by Theorem \ref{naima} this duadic code gives a
quantum duadic code $[[n,1,d]]_q$, which is impure as $d' \leq
\min\{p_i^{z_i}\}< \sqrt{n}<d.$
\end{proof}

{\small

\end{document}